\documentclass[twocolumn,times]{article}
\usepackage{mathptmx}
\usepackage[T1]{fontenc}
\usepackage{subcaption}
\usepackage[paper=a4paper]{geometry}

\usepackage{xspace}
\usepackage[final]{microtype}
\usepackage{hyperref}

\usepackage{graphicx}
\usepackage{float}
\usepackage{parskip}
\usepackage{textcomp}
\usepackage{anyfontsize}  

\usepackage[labelfont=bf]{caption}
\usepackage{xcolor,colortbl}
\usepackage{array, makecell}
\usepackage{multirow}
\usepackage{wrapfig}
\usepackage{hyperref}
\hypersetup{
    colorlinks=true,
    linkcolor=blue,
    filecolor=magenta,      
    urlcolor=black,
}

\advance\topmargin-1in\advance\oddsidemargin-1in
\evensidemargin\oddsidemargin

\makeatletter
\def\@startsection#1#2#3#4#5#6{\if@noskipsec \leavevmode \fi
   \par \@tempskipa #4\relax
   \@afterindenttrue
   \ifdim \@tempskipa <\z@ \@tempskipa -\@tempskipa \@afterindentfalse\fi
   \if@nobreak \everypar{}\else
     \addpenalty{\@secpenalty}\addvspace{\@tempskipa}\fi \@ifstar
     {\@dblarg{\@sect{#1}{#2}{#3}{#4}{#5}{#6}}}%
     {\@dblarg{\@sect{#1}{#2}{#3}{#4}{#5}{#6}}}}

\def\section{\@startsection {section}{1}{\z@}{1pt plus 1pt minus 1pt}
{1pt plus 1pt minus 1pt}{\centering\normalsize\bf}}

\long\def\@makecaption#1#2{
\vskip10pt\begin{center} #1 #2 \end{center}\par\vskip 1pt}
\def\fnum@figure{\raggedright{\small Fig. \thefigure }.%
\small}
\def\fnum@table{\small TABLE \thetable\\\small}
\def\thetable{\Roman{table}}

\def\thebibliography#1{\section*{References\@mkboth
 {REFERENCES}{REFERENCES}}\list
 {[\arabic{enumi}]}{\settowidth\labelwidth{[#1]}\leftmargin\labelwidth
 \parsep 0pt \itemsep 2pt plus 1pt minus 1pt
 \advance\leftmargin\labelsep
 \usecounter{enumi}}
 \def\newblock{\hskip .11em plus .33em minus .07em}
 \sloppy\clubpenalty4000\widowpenalty4000
 \sfcode`\.=1000\relax}

\makeatother

\colorlet{myGray}{gray!40}

\usepackage{tikz}

\usepackage{titlesec}

\titlespacing\section{0pt}{0pt plus 3pt minus 3pt}{0pt plus 3pt minus 3pt}

\date{}

\title{
\large\bf				       
Clo-HDnn: A 4.66 TFLOPS/W and 3.78 TOPS/W  Continual On-Device Learning Accelerator \\ with  Energy-efficient Hyperdimensional Computing via Progressive Search \vspace{-4mm}}

\author{\normalsize Chang Eun Song*$^1$, Weihong Xu*$^1$, Keming Fan$^1$, Soumil Jain$^1$, Gopabandhu Hota$^1$, Haichao Yang$^1$,\
Leo Liu$^2$, \\ \normalsize Kerem Akarvardar$^2$, Meng-Fan Chang$^2$, Carlos H. Diaz$^2$,  Gert Cauwenberghs$^1$, Tajana Rosing$^1$, and Mingu Kang$^1$ (* co-first) \\
\normalsize $^1$University of California San Diego, La Jolla, CA, USA\
\normalsize $^2$Taiwan Semiconductor Manufacturing Company (TSMC)\vspace{-6mm}}

\begin{document}
\fontsize{10}{10}\selectfont
\maketitle

\thispagestyle{plain}
\pagenumbering{arabic}

\thispagestyle{empty}

\noindent \textbf{Abstract:} Clo-HDnn is an on-device learning (ODL) accelerator designed for emerging continual learning (CL) tasks. Clo-HDnn integrates hyperdimensional computing (HDC) along with low-cost Kronecker HD Encoder and weight clustering feature extraction (WCFE) to optimize accuracy and efficiency. Clo-HDnn adopts gradient-free CL  to efficiently update and store the learned knowledge in the form of class hypervectors. 
Its dual-mode operation enables bypassing costly feature extraction for simpler datasets, while progressive search reduces complexity by up to 61\% by encoding and comparing only partial query hypervectors. Achieving 4.66 TFLOPS/W (FE) and 3.78 TOPS/W (classifier), Clo-HDnn delivers 7.77$\times$ and 4.85$\times$ higher energy efficiency compared to SOTA ODL accelerators.

\noindent \textbf{Introduction:} 
Continual learning (CL) is a framework that mimics human learning, allowing models to adapt to new tasks or evolving data (Fig.1). 
This capability is crucial for various emerging applications exposed to dynamic environments, where obtained data patterns change over time.
The CL on edge devices require a low-power, compact solution that preserves prior knowledge while continually learning and adapting [1,2].
However, existing ODL accelerators face following key challenges: \textbf{(C1)} Gradient-based training is computationally expensive 
due to the complex data flow and float-point operations~[3].
\textbf{(C2)} Furthermore, during training, new data often overwrites prior knowledge, limiting long-term CL capability. Hyperdimensional computing (HDC), a brain-inspired computational framework that operates on high-dimensional vectors, offers alternative light-weight CL solution. The independence of class hypervectors (CHVs) makes HDC well-suited for dynamic and incremental learning (Fig.2). However, traditional HDC accelerators present additional challenges. \textbf{(C3)} Large encoders and exhaustive distance searches across multiple HVs increase computational overhead [13] while storing CHVs demands a significant associative memory (AM) capacity [4].

To address these issues, this paper introduces Clo-HDnn with following features (Fig.3): \textbf{(S1)} Gradient-free HDC-based training combined with an efficient feature extractor (FE) reduces computational costs. A dual-mode operation enables bypassing FE for certain datasets, further optimizing computation costs. \textbf{(S2)} HDC enables knowledge retention while adapting to new tasks, preventing catastrophic forgetting. \textbf{(S3)} A Kronecker HD encoder [13] and progressive search accelerate inference with fewer computations, reducing cache usage and computing complexity. Additionally, Clo-HDnn incorporates a customized instruction set architecture (ISA) to enhance programmability.

\noindent \textbf{Proposed Design:} Fig.3 shows the Clo-HDnn overview, composed of the WCFE, HD module connected by the global FIFO module. The WCFE extracts features using weight clustering, while the HD module performs encoding, training, and inference. 
Fig.4 shows the dual-mode data flows supported by the flexible CDC FIFO connection in Clo-HDnn. First, the dual-mode operation allows simple datasets to bypass the WCFE during inference, reducing computation, while complex datasets are processed via the WCFE before the HD classifier. Second, progressive search only encodes the input feature into a partial segment of the query hypervector (QHV) and compares it with the associative partial CHVs. If the distance margin between classes exceeds the preset confidence threshold, the remaining encoding and search processes are terminated, reducing complexity by up to 61\% with negligible accuracy loss.

Fig.5 shows the proposed Kronecker HD Encoder that exploits Kronecker product to significantly optimize the computation \& memory overhead, as encoding is the main bottleneck in HDC. It comprises two stages with reshaping and block matrix multiplications. To support progressive search, the encoder generates partial QHVs by computing only the necessary block of each matrix with the input features, further reducing computation and latency. With an 8-bank 1KB weight buffer processing 256-b weights per cycle and segmented feature vectors streaming into 32 8-to-1 adder trees, the binary-INT encoding reduces multiplication to addition, achieving 43$\times$ speedup, 1376$\times$ memory capacity savings compared to existing lengthy encoding methods (RP [11], cRP [4], and ID-LEVEL (ID) [12]).

Fig.6 illustrates the data flow of HDC. \textcircled{\raisebox{-1.5pt}{1}} Input features are encoded into QHVs using the Kronecker HD Encoder while CHVs are stored in cache, with each column containing INT8 elements. During inference, \textcircled{\raisebox{-1.5pt}{2}} progressive search generates a partial QHV. For each cycle, the 64-b MSBs of each CHV are fetched and compared against the segmented QHV using the XOR tree in HD Search module, identifying the best match. A key advantage of progressive search is that only partial CHVs need to be stored in the cache, significantly reducing cache size and bandwidth requirements. The HDC Training module supports single-pass training and retraining, where CHVs are updated by adding or subtracting QHV based on inference correctness and written back to CHV cache memory.

Fig.7 details the WCFE for efficient feature extraction using the following techniques: (a) After training, pretrained weights with similar values are clustered, and each filter weight is indexed via weight codebook. (b) During multiplication, inputs sharing the same weights are grouped, accumulated and multiplied once for efficient processing. (c) A $4\times16$ Processing Element (PE) array with 4 register files (RFs) and 1 MAC per PE enables parallel accumulation and multiplication, minimizing idle time. This reduces parameters by 1.9$\times$ and CONV computations by 2.1$\times$.

\noindent \textbf{ISA and Programming Model:} Fig.8 depicts the ISA and execution model, enabling programmability for WCFE, HDC, and global FIFO with two types of customized instructions for memory and arithmetic, using a unified 20-bit instruction format (4-b opcode \& 16-b operand). We implement C/C++ intrinsics for each instruction to provide an interface to high-level CL application code. The inline assembly operator is compiled and used to emit the bytecode of corresponding instructions. 

\noindent \textbf{Performance Results:} 
Fig.9 presents the end-to-end CL accuracy across three benchmarks: CIFAR100 [7] in normal mode, ISOLET [14] \& UCIHAR [15] in bypassing mode. Clo-HDnn achieved negligible accuracy drop as the FP baseline [5]. Fig.10 highlights the measured performance, showing that the WCFE and HDC modules operate at 50-250 MHz across 0.7-1.2V. Energy efficiency reaches 1.44-4.66 TFLOPS/W for the WCFE and 1.29-3.78 TOPS/W for the HDC. The breakdown reveals that the WCFE accounts for 94.2\% of total energy consumption and 87.7\% of the latency, which can be effectively reduced by WCFE bypassing. Fig.11 shows the $14.4$~mm$^2$ chip micrograph, implemented in 40~nm CMOS. The comparison table confirms that Clo-HDnn is the first chip to support end-to-end CL for HDC tasks, delivering 1.73-7.77$\times$ and 4.85$\times$ higher energy efficiency for the CNN (WCFE) and classifier (HDC) compared to state-of-the-art accelerators [4, 8].

\clearpage

\begin{figure}[!ht]
\centering
\begin{tikzpicture}
\draw[thick] (0,0) rectangle (18,25);
\foreach \x in {9} {
    \draw[thick] (\x,0) -- (\x,25);
}
\foreach \y in {25/3,25/3*2} {
    \draw[thick] (0,\y) -- (18,\y);
}

\node[anchor=center] at (4.5,21.3) {\includegraphics[width=1\linewidth]{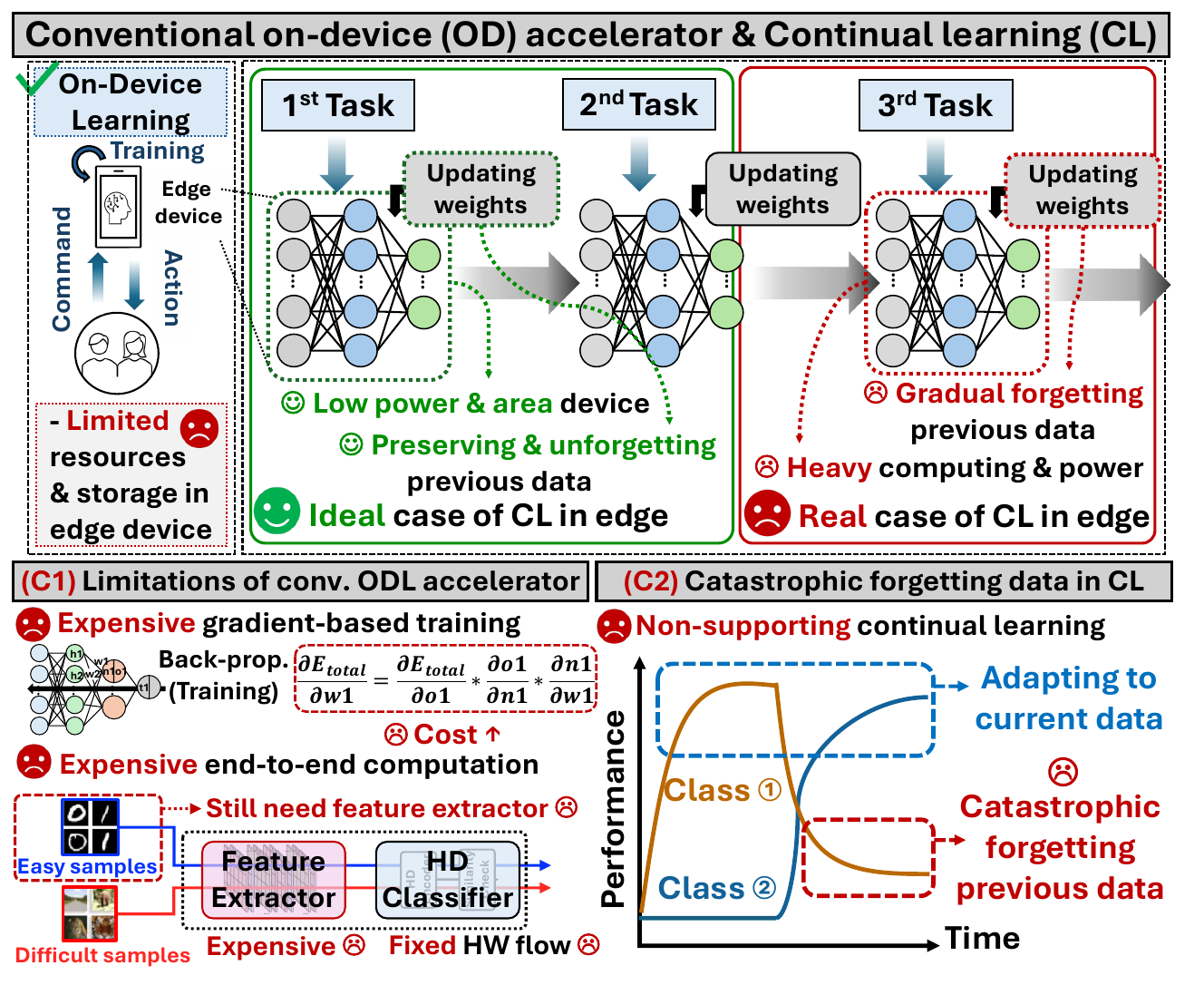}};
\node[anchor=center, text width=1\linewidth] at (4.5,17.2) {\fontsize{9}{8}\selectfont Fig.1. Overview of conventional on-device accelerator, continual learning (CL), and their challenges.};

\node[anchor=center] at (13.5,21.3) {\includegraphics[width=1\linewidth]{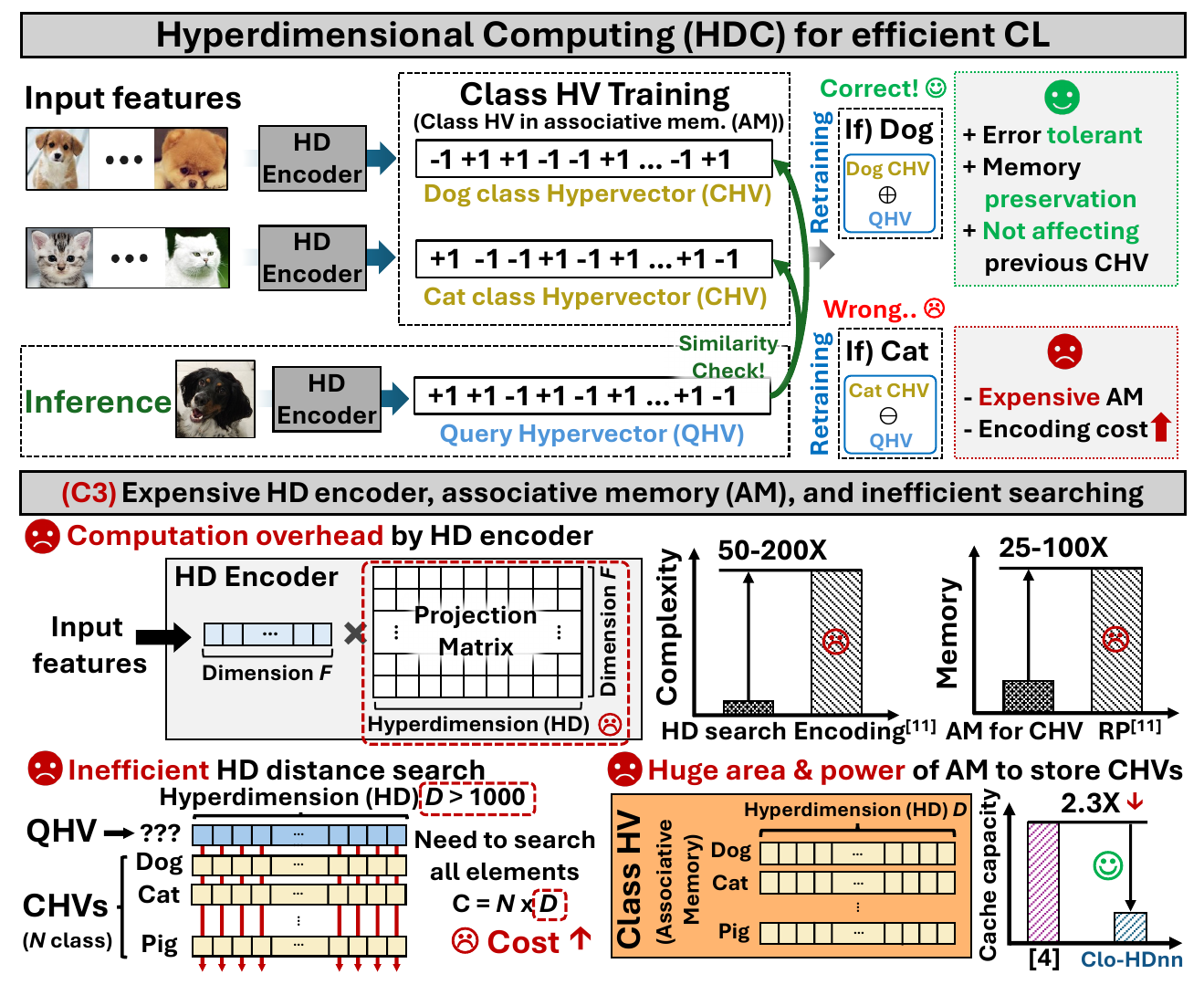}};
\node[anchor=center, text width=1\linewidth] at (13.5,17.2) {\fontsize{9}{8}\selectfont Fig.2. Data flow of hyperdimensional computing (HDC) and its benefits and challenges for CL.};

\node[anchor=center] at (4.5,12.93) {\includegraphics[width=1\linewidth]{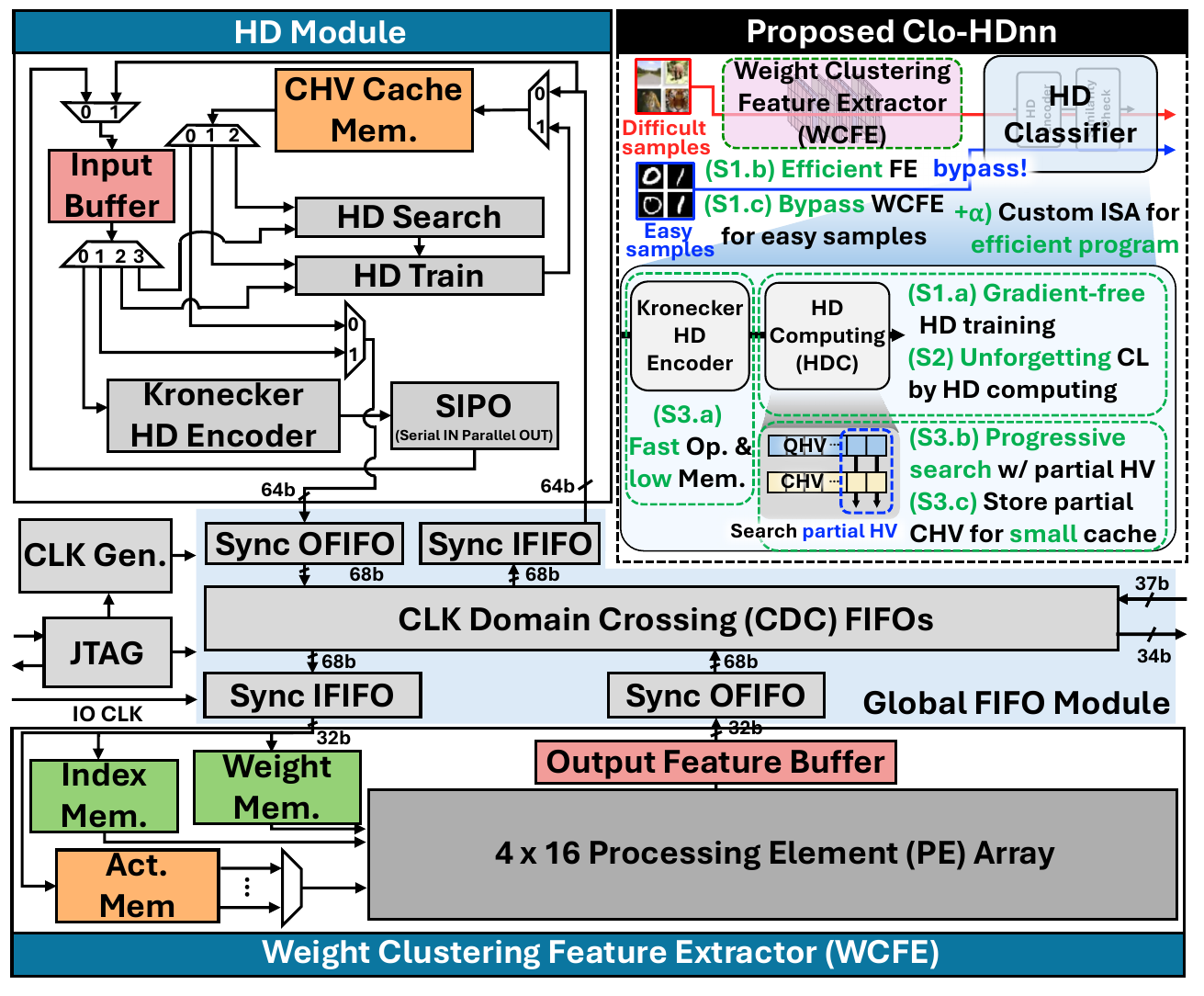}};
\node[anchor=center, text width=1\linewidth] at (4.5,8.8) {\fontsize{9}{8}\selectfont Fig.3. Proposed Clo-HDnn  architecture and key features.};

\node[anchor=center] at (13.5,12.93) 
{\includegraphics[width=1\linewidth]{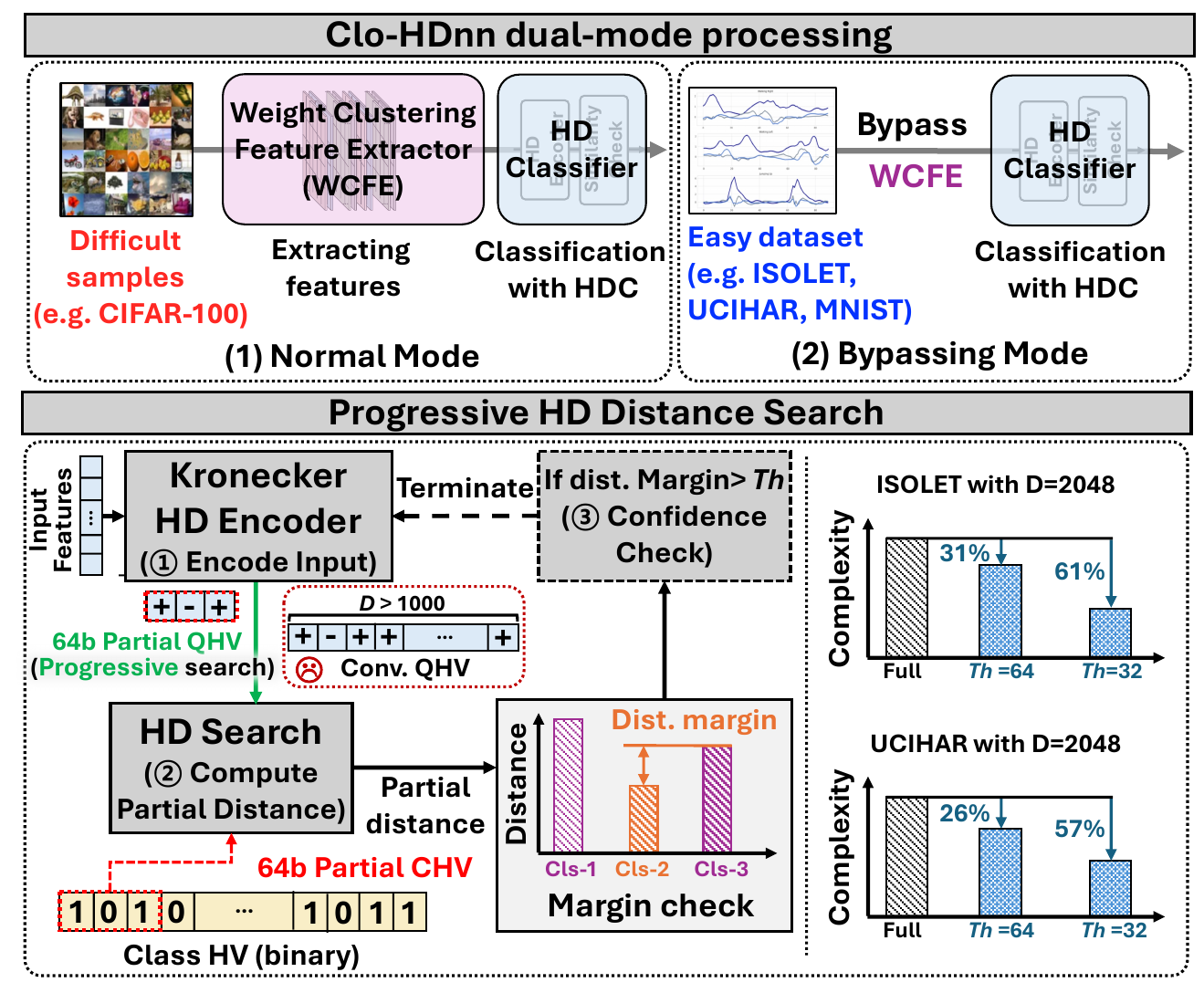}};
\node[anchor=center, text width=1\linewidth] at (13.5,8.8) {\fontsize{9}{8}\selectfont Fig.4. Data flow of Clo-HDnn's dual-mode processing and proposed progressive HD distance search.};

\node[anchor=center] at (4.5,4.5) 
{\includegraphics[width=1\linewidth]{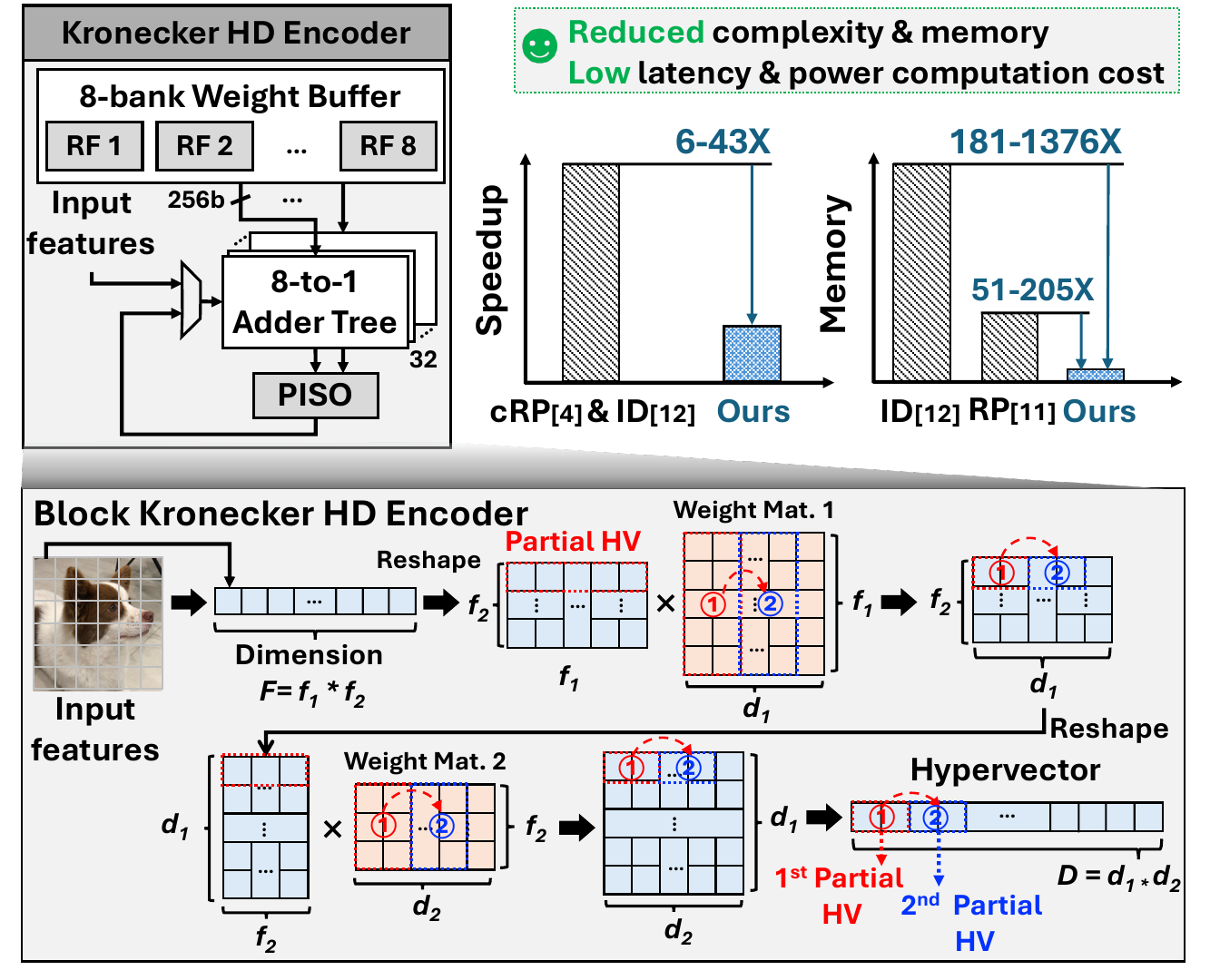}};
\node[anchor=center, text width=1\linewidth] at (4.5,0.48) {\fontsize{9}{8}\selectfont Fig.5. Details of proposed kronecker HD encoder and its benefits compared to cyclic random projection (cRP) and conventional RP.};

\node[anchor=center] at (13.5,4.5) 
{\includegraphics[width=1\linewidth]{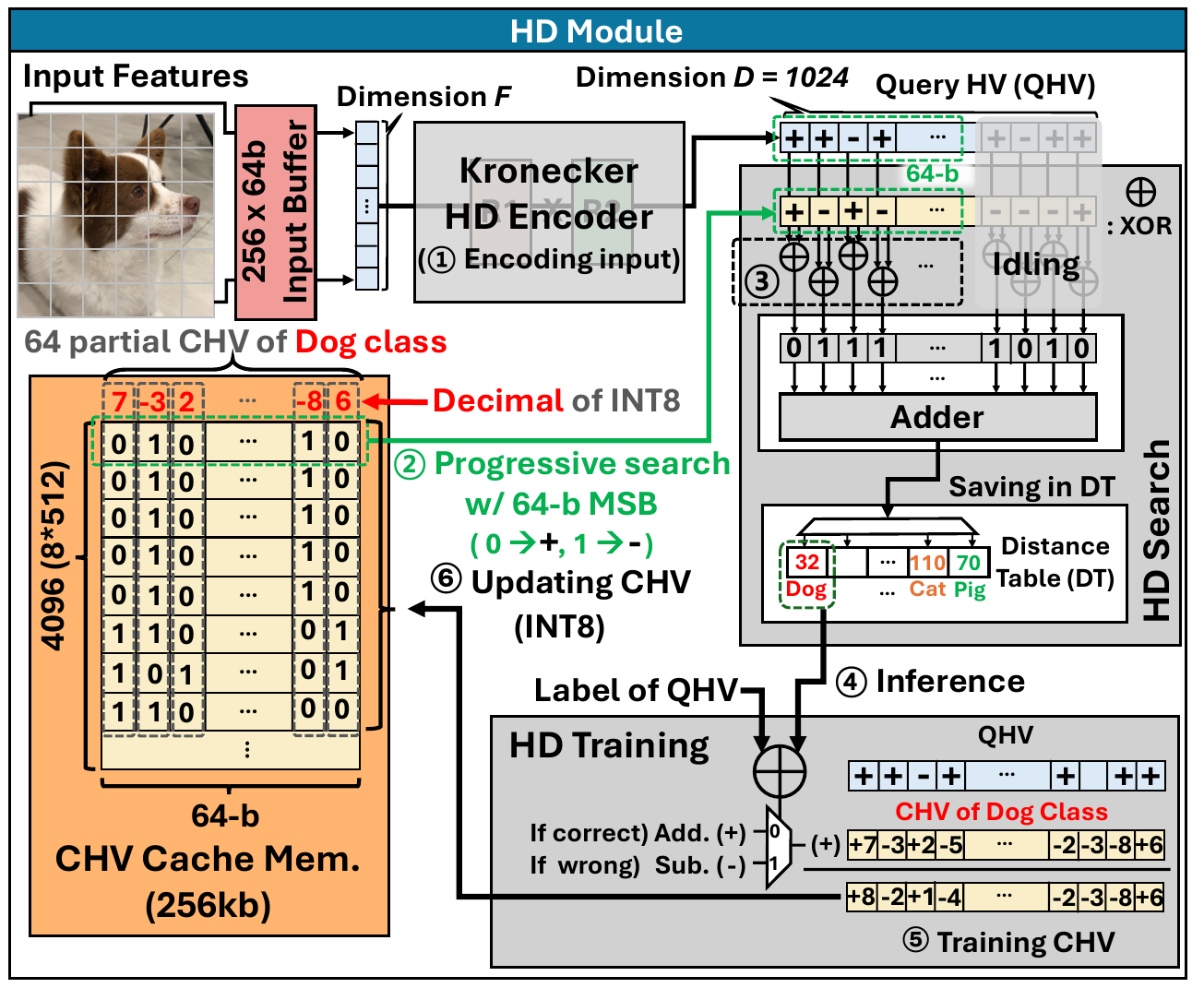}};
\node[anchor=center, text width=1\linewidth] at (13.5,0.48) {\fontsize{9}{9}\selectfont Fig.6. Dataflow of inference and training  in HD module with progressive search mechanism.};

\end{tikzpicture}
\end{figure}

\clearpage

\begin{figure}[!ht]
\centering
\begin{tikzpicture}
\draw[thick] (0,0) rectangle (18,25);
\foreach \x in {9} {
    \draw[thick] (\x,0) -- (\x,25);
}
\foreach \y in {25/3,25/3*2} {
    \draw[thick] (0,\y) -- (18,\y);
}

\node[anchor=center] at (4.5,21.2) {\includegraphics[width=1\linewidth]{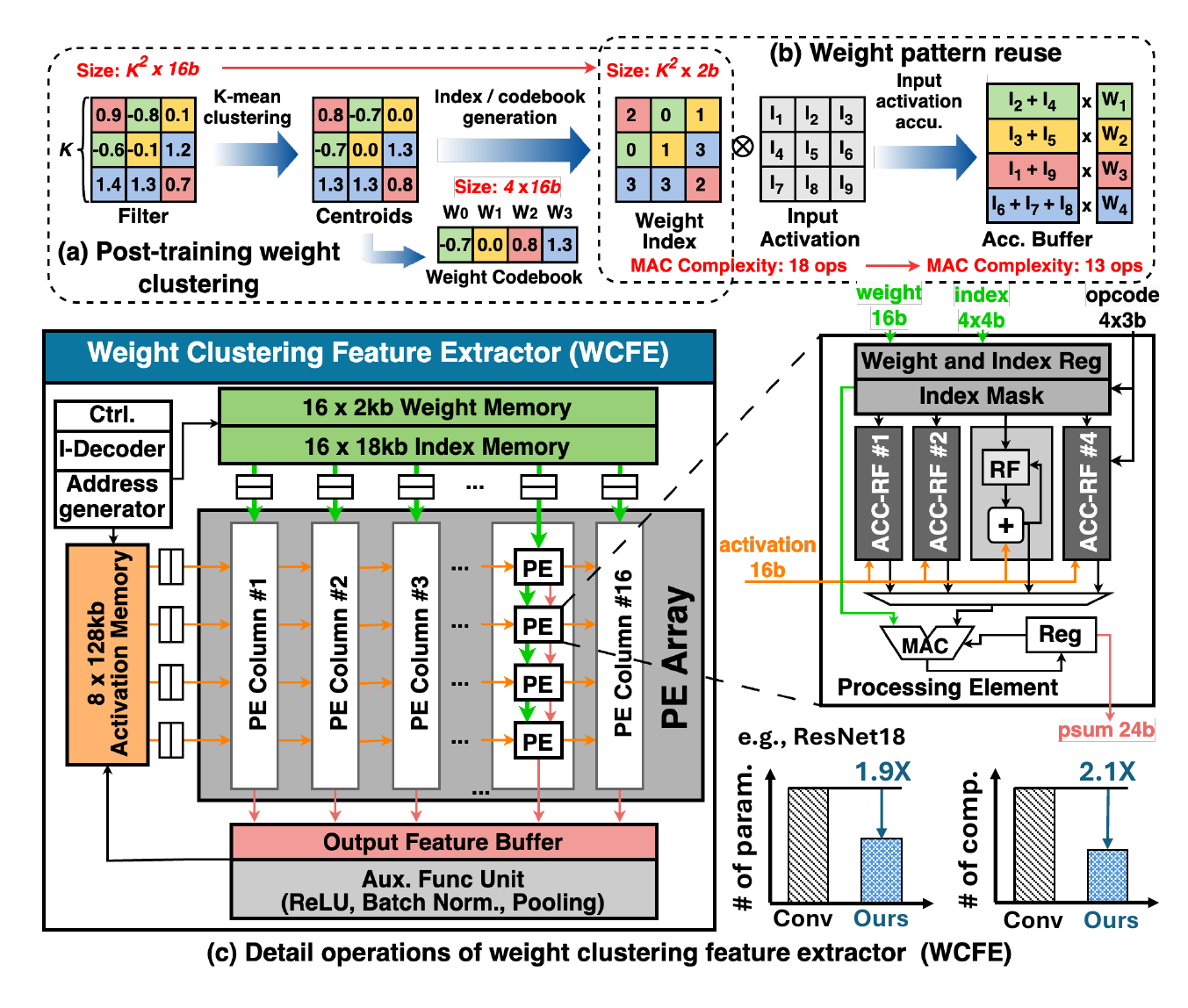}};
\node[anchor=center, text width=1\linewidth] at (4.5,17.2) {\fontsize{9}{8}\selectfont Fig.7. (a) Post-training weight clustering of filters, (b) pattern reusing during inference, and (c) WCFE architecture.};

\node[anchor=center] at (13.5,21.3) {\includegraphics[width=1\linewidth]{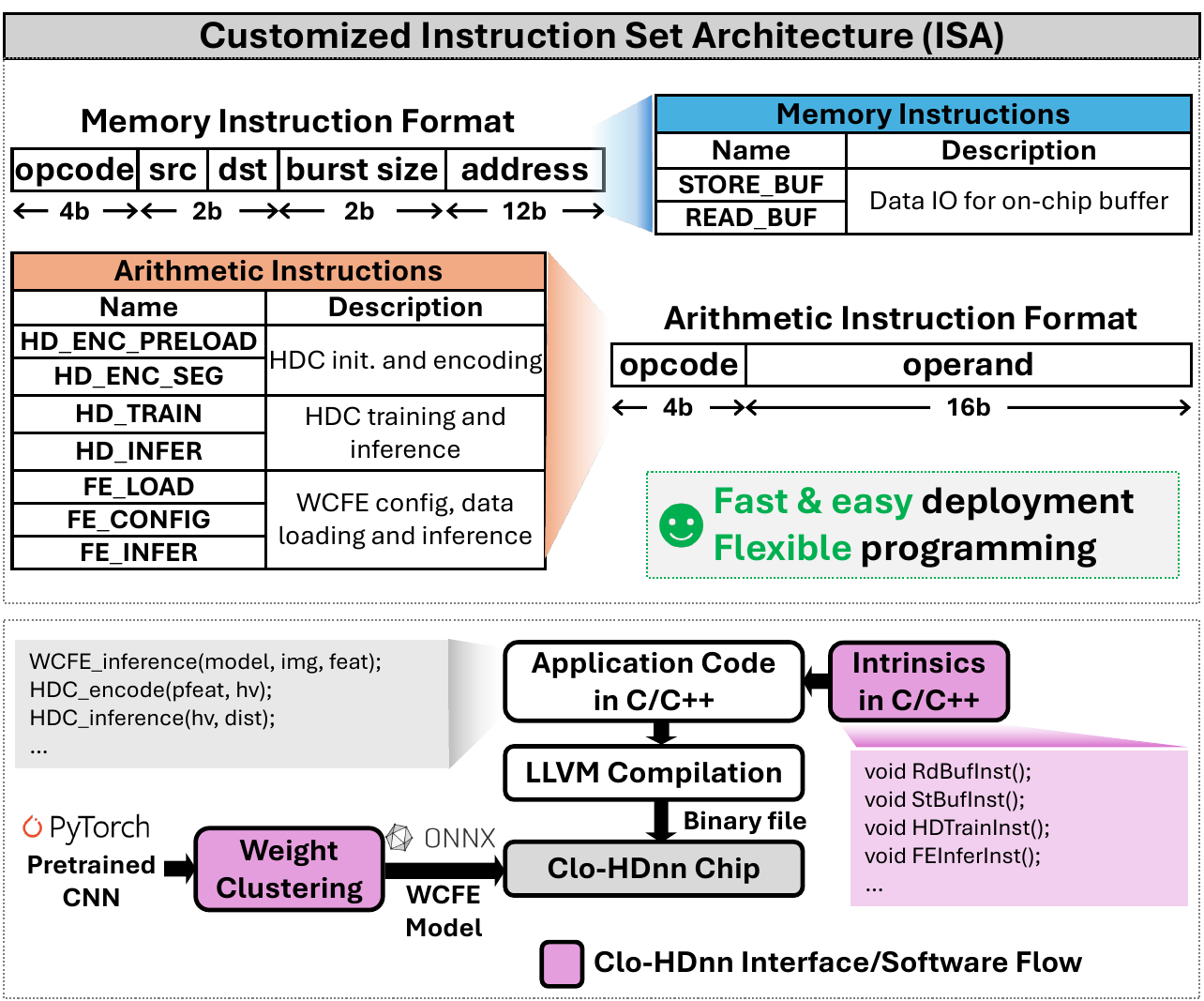}};
\node[anchor=center, text width=1\linewidth] at (13.5,17.2) {\fontsize{9}{8}\selectfont Fig.8. Customized instruction set architecture (ISA) and programming model of Clo-HDnn chip.};

\node[anchor=center] at (4.5,12.93) {\includegraphics[width=1\linewidth]{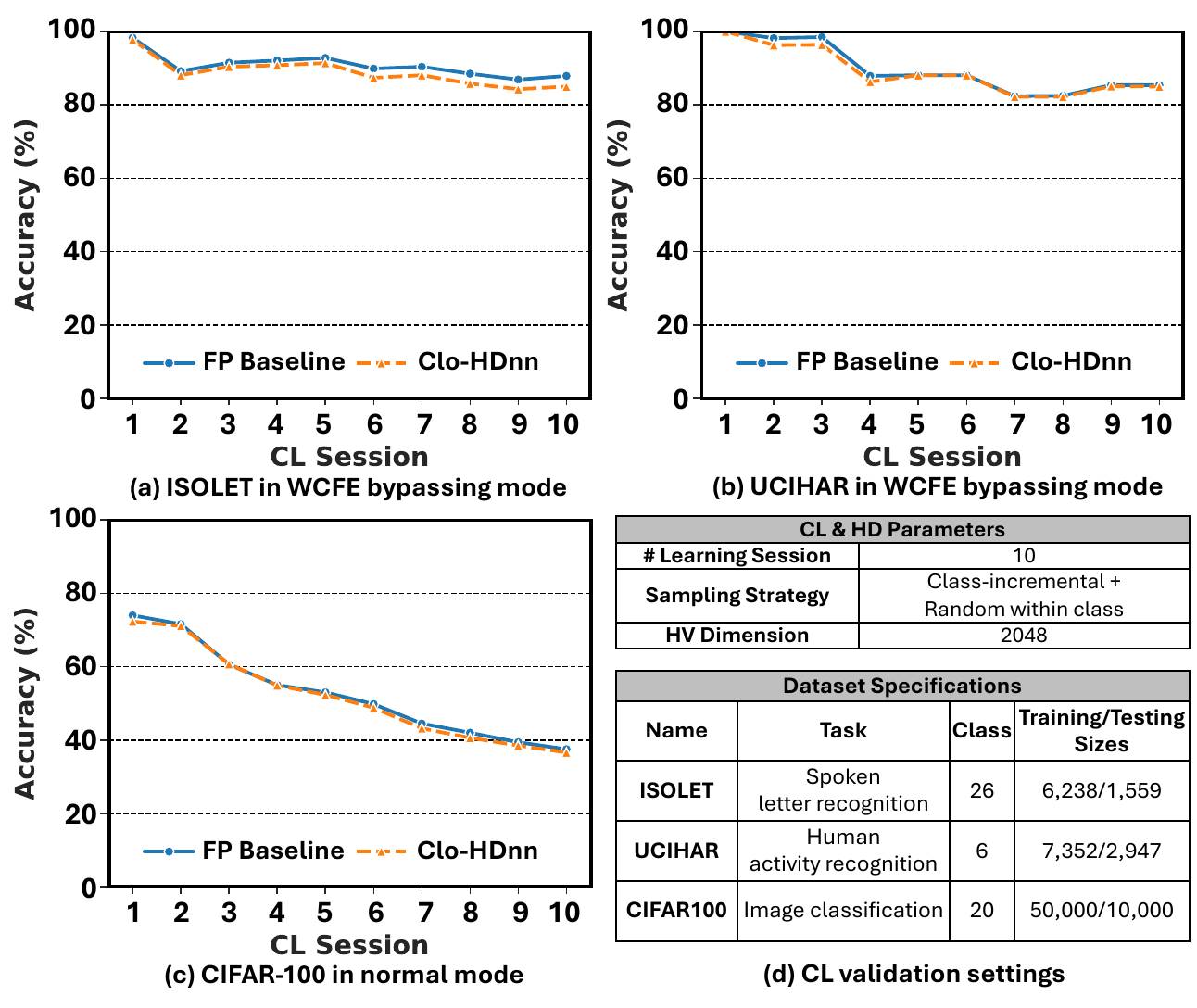}};
\node[anchor=center, text width=1\linewidth] at (4.5,8.8) {\fontsize{9}{8}\selectfont Fig.9. Accuracy results of bypassing WCFE ((a) ISOLET \& (b) UCIHAR) and normal mode ((c) CIFAR-100) compared to [5].};

\node[anchor=center] at (13.5,12.9) {\includegraphics[width=1\linewidth]{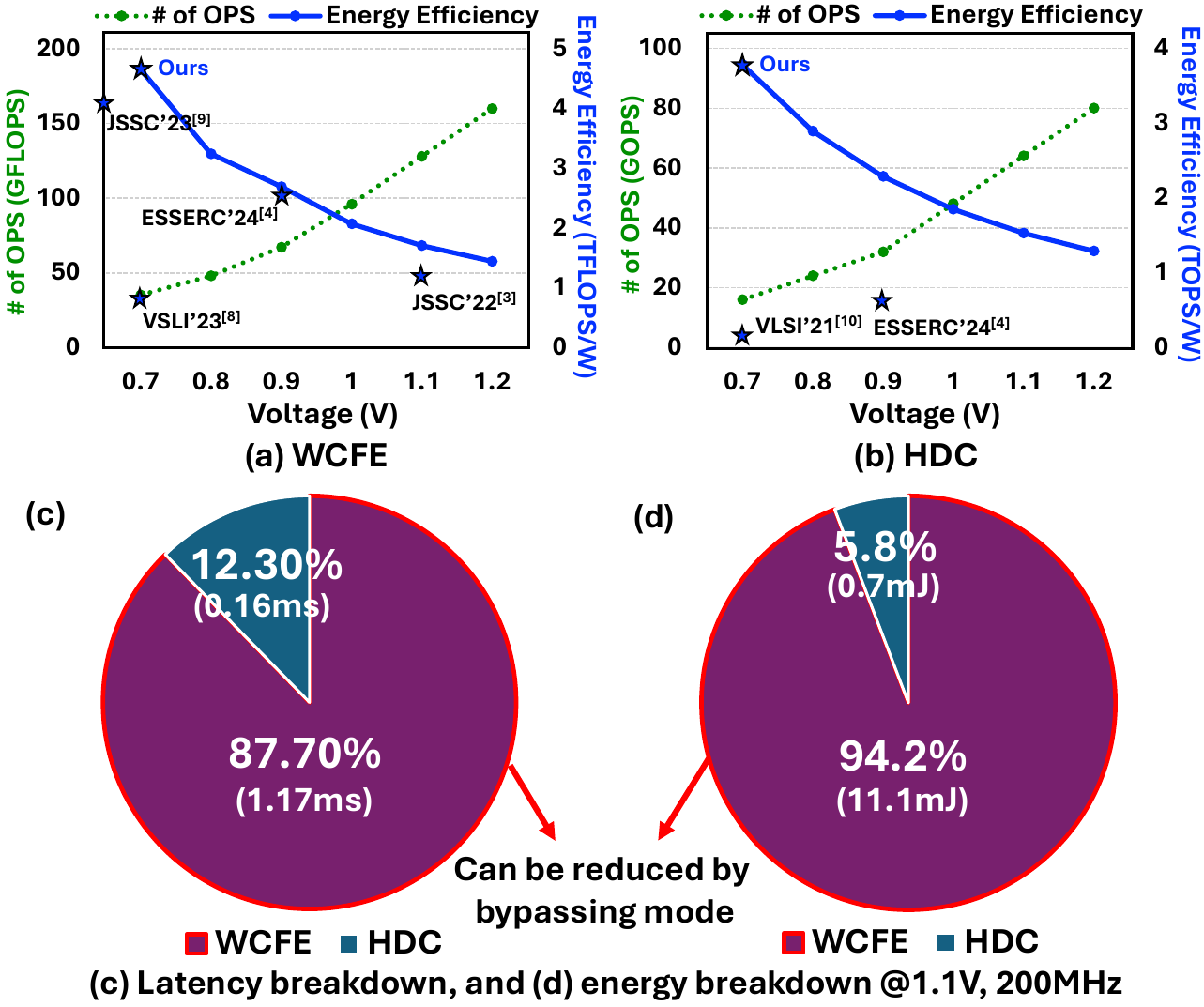}};
\node[anchor=center, text width=1\linewidth] at (13.5,8.8) {\fontsize{9}{8}\selectfont Fig.10. Energy efficiency and peak throughput of (a) WCFE, (b) HDC, and (c) latency,  (d) energy breakdowns of CIFAR-100.};

\node[anchor=center] at (4.5,4.5) {\includegraphics[width=1\linewidth]{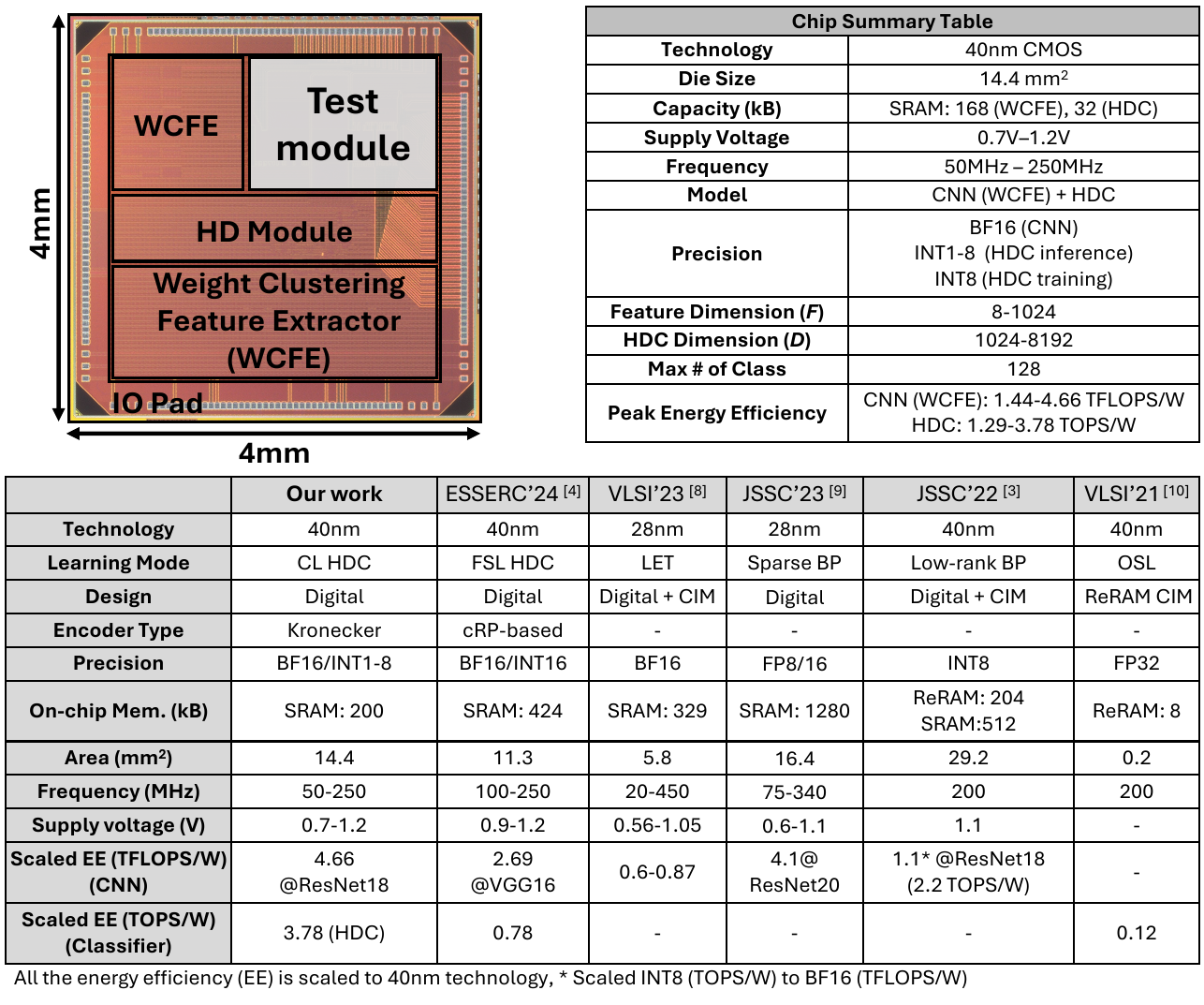}};
\node[anchor=center, text width=1\linewidth] at (4.5,0.48) {\fontsize{9}{8}\selectfont Fig.11. Chip die photo, Clo-HDnn summary table, and comparison table with SOTA.};

\node[anchor=north] at (13.5, 8) {%
    \fontsize{9}{10}\selectfont
    \textbf{Acknowledgement}
};

\node[anchor=north, text width=1\linewidth, align=left] at (13.5, 7.5) {%
    \fontsize{9}{10}\selectfont
    This work was supported by TSMC and in part by PRISM and CoCoSys, centers in JUMP 2.0, an SRC program sponsored by DARPA. \#449008
};

\node[anchor=north] at (13.5, 6.2) {%
    \fontsize{9}{10}\selectfont
    \textbf{References}
};

\node[anchor=north, text width=1\linewidth, align=left] at (13.5, 5.7) {%
    \fontsize{9}{10}\selectfont
    [1] X. Ma et al., MobiCom, 2023, Article 83, 1-15. \\[-0.05em] 
    [2] G. Karunaratne et al., ESSCIRC, 2022, pp. 105-108. \\[-0.05em]
    [3] K. Prabhu, et al., JSSC, vol. 57, no. 4, pp. 1013-1026, 2022. \\[-0.05em]
    [4] H. Yang et al., ESSERC, 2024, pp. 33-36. \\[-0.05em]
    [5] X. Yu et al., IPSN, 2024. \\[-0.05em]
    [6] I. Kazi et al., TCAS-I, vol. 61, no. 11, pp. 3155-3164, 2014. \\[-0.05em]
    [7] A. Krizhevsky et al., (2009): 7.\\[-0.05em]
    [8] J. -H. Kim, et al., Symp. on VLSI, 2023, pp. 1-2. \\[-0.05em]
    [9] S. K. Venkataramanaiah et al., JSSC, vol. 58, no. 7, 2023. \\[-0.05em]
    [10] H. Li et al., Symp. on VLSI, 2021, pp. 1-2. \\[-0.05em]
    [11] A. Hernandez-Cano et al., DAC, 2021. p. 7-12. \\[-0.05em]
    [12] M. Imani et al., ICRC 2017 (pp. 1-8). IEEE. \\[-0.05em]
    [13] W. Xu et al., ICCD, 2023, pp. 243-246. \\[-0.05em]
    [14] UCI machine learning repository. ISOLET. \\[-0.05em]
    [15] J. Reyes-Ortiz et al., UCI ML Repository, 2013. \\[-0.05em]
};

\end{tikzpicture}
\end{figure}


\end{document}